\documentclass{PoS}

\def\321{$\mathrm{SU(3) \otimes SU(2) \otimes U(1)}$ }
\def\lfv{lepton flavor violation }

\def \znbb {$\rm 0\nu\beta\beta$ }

\def\gsim{\raise0.3ex\hbox{$\;>$\kern-0.75em\raise-1.1ex\hbox{$\sim\;$}}}
\def\lsim{\raise0.3ex\hbox{$\;<$\kern-0.75em\raise-1.1ex\hbox{$\sim\;$}}}
\def\lfv{lepton flavor violation }

\def\SM{$\mathrm{SU(3)_c \otimes SU(2)_L \otimes U(1)_Y}$ }
\newcommand{\sm}{{Standard Model }}

\def\gsim{\raise0.3ex\hbox{$\;>$\kern-0.75em\raise-1.1ex\hbox{$\sim\;$}}}
\def\lsim{\raise0.3ex\hbox{$\;<$\kern-0.75em\raise-1.1ex\hbox{$\sim\;$}}}

\definecolor{linkcolor}{rgb}{0,0,0.5}

 \usepackage[usenames,dvipsnames]{xcolor}
 \usepackage[normalem]{ulem}

\title{Neutrino oscillations and flavor theories }

\ShortTitle{Neutrino oscillations and flavor theories }

\author{{Jos\'e W. F. Valle}
\thanks{ This work was supported by the Spanish grants FPA2017-85216-P (AEI/FEDER, UE), PROMETEO/2018/165 (Generalitat Valenciana) and the Spanish Red Consolider MultiDark FPA2017-90566-REDC. }\\
\textcolor{black}{AHEP Group, Institut de F\'{i}sica Corpuscular --   C.S.I.C.\\
Universitat de Val\`{e}ncia, Parc Cientific de Paterna.\\   C/Catedratico Jos\'e Beltr\'an, 2 E-46980 Paterna (Val\`{e}ncia) - SPAIN}\\
        E-mail: \email{valle@ific.uv.es}}

      \abstract{I discuss neutrino mixing ansatze, such as the generalized Tri-bimaximal and bi-large mixing patterns, and their utility in describing the oscillation data.
       Unitarity tests and probes of the absolute neutrino mass scale are briefly discussed.
       A short overview of neutrino mass generation is given.
       I discuss an orbifold approach to the flavor problem and the resulting implications, e.g. the golden quark-lepton mass relation, \znbb and neutrino oscillation predictions.
}

\FullConference{Proceedings of the 40th International Conference on High Energy Physics (ICHEP2020)\\
  July 28 - August 6, 2020,
  Prague,  Czech Republic\\
}

\begin{document}

\begin{figure}[h]
    \centering \vglue -.4cm
    \includegraphics[height=4cm,width=0.35\textwidth]{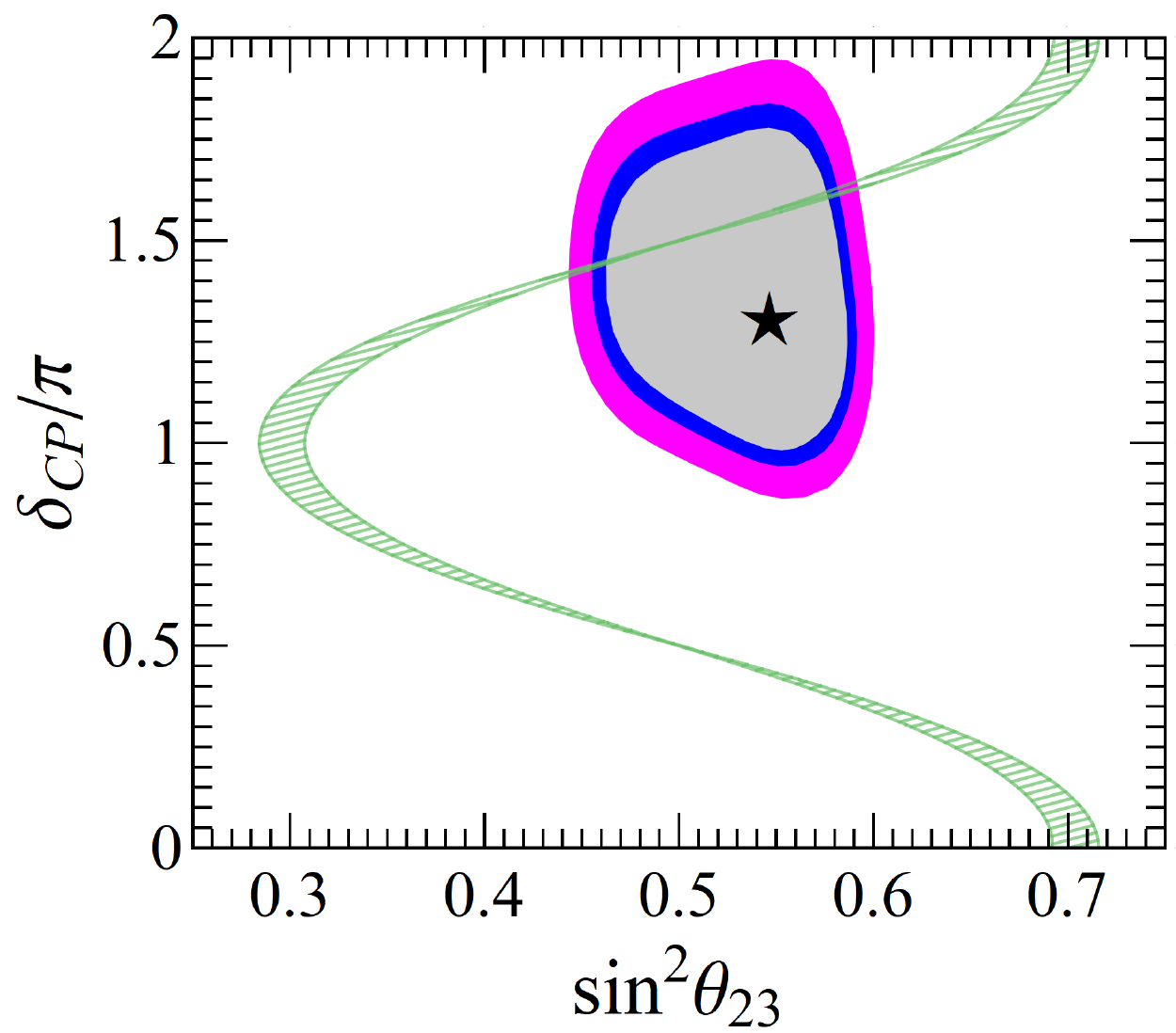}~~~~~
    \includegraphics[height=4cm,width=0.35\textwidth]{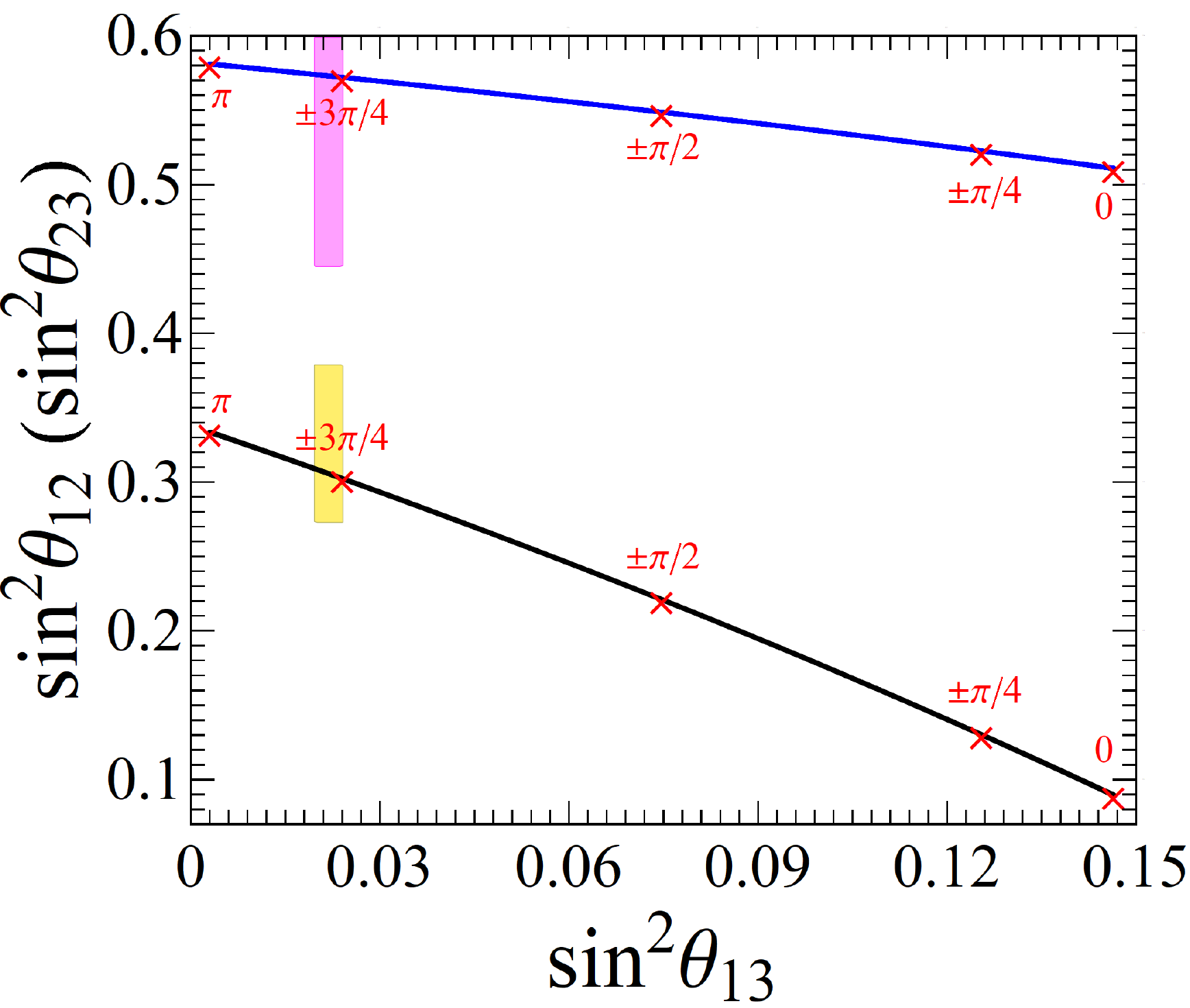}
    \vglue -.4cm
    \caption{Left: Predicting the CP phase in a realistic generalized TBM scenario, from~\cite{Chen:2018eou}.
      Right: Predicting solar and atmospheric mixing angles within the ``bi-large'' scheme of Ref.~\cite{Chen:2019egu}.} 
    \label{fig:tbm}
  \end{figure}

{\bf Neutrino mixing }~
Here I start with the interpretation of the neutrino oscillation results reported by Tortola.
A striking feature of the observed pattern of neutrino mixing, is that it is characterized by two large angles, one nearly maximal and the other nearly thirty degrees.
A simple way to account for this is the TBM ansatz.
Although it captures the most salient features of the oscillation phenomenon, TBM is now in conflict with the non-zero value of $\theta_{13}$ indicated by reactor experiments~\cite{An:2012eh}.
Fortunately, there are systematic ways of generalizing patterns with manifest $\mu-\tau$ reflection symmetry, such as TBM.
By exploiting partially conserved generalized CP symmetries~\cite{Chen:2015siy,Chen:2016ica,Chen:2018lsv} one can develop ``revamped'' TBM patterns.
This way one obtains two-parameter mixing schemes which can be not only viable, but also predictive~\cite{Nath:2018fvw}.
As an example, we present one in which the CP phase is predicted to lie on the green band in Fig.~\ref{fig:tbm}, left~\cite{Chen:2018eou}. 
Clearly the overlap of the predicted band with the 3-neutrino global oscillation region covers a narrow $\delta_{CP}$ range~\footnote{This scheme also promotes the precise measurement of $\theta_{13}$ into predictions for the solar angle.}.

  Bi-large neutrino mixing~\cite{Boucenna:2012xb,Ding:2012wh} builds up on the observation that the Cabibbo angle is similar in magnitude to the reactor mixing angle and may be used to seed flavor mixing.
One finds that the good measurement of $\theta_{13}$ at reactors allows us to predict \textit{both} solar and atmospheric mixing angles~\cite{Chen:2019egu} quite acurately,
as seen in the right panel of Fig.~\ref{fig:tbm}.
It also leads to sharp predictions for the modulus of the CP phase.
There are also ``softer'' bi-large-type lepton mixing schemes in which the oscillation parameters are given in terms of two free parameters~\cite{Roy:2014nua}. \\[-.4cm]

{\bf Unitarity tests}~ 
  If light neutrinos acquire their masses from the exchange of heavy neutral singlet mediators, as in the simplest seesaw mechanism, 
  then their propagation will be described effectively by a non-unitary mixing matrix~\cite{Valle:1987gv,Nunokawa:1996tg,Antusch:2006vwa,Escrihuela:2015wra,Miranda:2016ptb}.
%  
% %
Most generally, this is expressed as 
\begin{equation}
\left(\begin{array}{ccc}
\alpha_{11} & 0 & 0\\
\alpha_{21} & \alpha_{22} & 0\\
\alpha_{31} & \alpha_{32} & \alpha_{33}
\end{array}\right)~U_{{\rm }},\label{eq:s-6}
\end{equation}
where the unitary matrix $U$ is pre-multiplied by a triangular matrix~\cite{Valle:1987gv,Escrihuela:2015wra,Miranda:2016ptb}
whose diagonal entries are real and less than unity, while the off-diagonal ones are small but complex. 
Due to an intrinsic degeneracy between $\delta_{CP}$ and the phase in $\alpha_{21}$~\cite{Miranda:2016wdr} it so happens that the latter
plays a crucial role in the CP measurement.
This confusion problem can be mitigated~\cite{Ge:2016xya} by having better limits on the magnitude of $\alpha_{21}$. 
Current bounds are at the \% level, see Ref.~\cite{Escrihuela:2016ube}.
The short-baseline neutrino program at Fermilab is ideally suited to probe the unitarity of the lepton mixing matrix, with potentially better sensitivities~\cite{Miranda:2018yym}. 
Apart from their motivation as a seesaw probe, unitarity tests at short distance setups are required for a robust CP measurement in the far detector.\\[-.4cm]
 
\begin{figure}[h]
    \centering \vglue -.4cm
\includegraphics[height=5.6cm,width=0.45\textwidth]{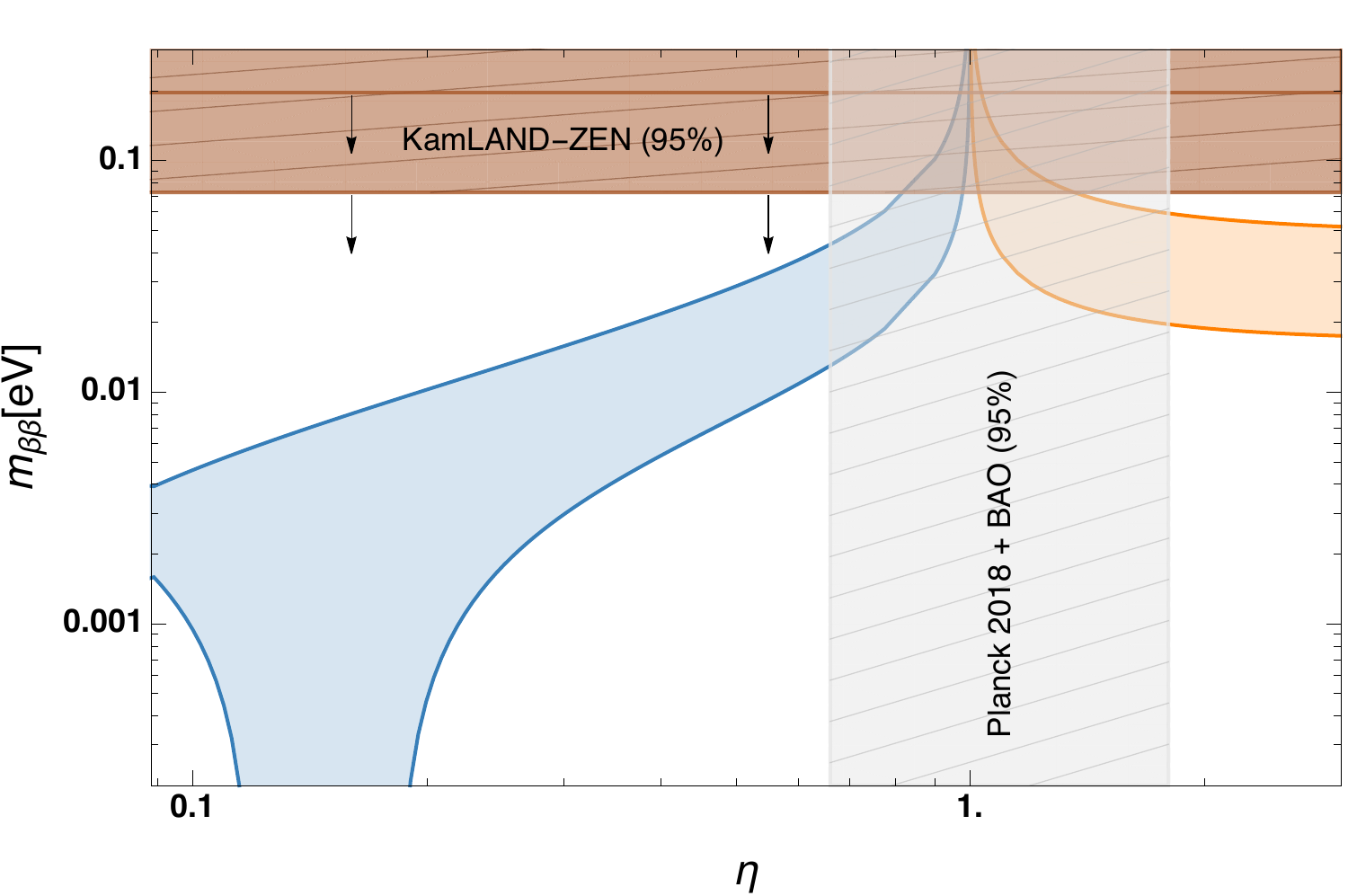}~~~~~%
\includegraphics[height=5.2cm,width=0.52\textwidth]{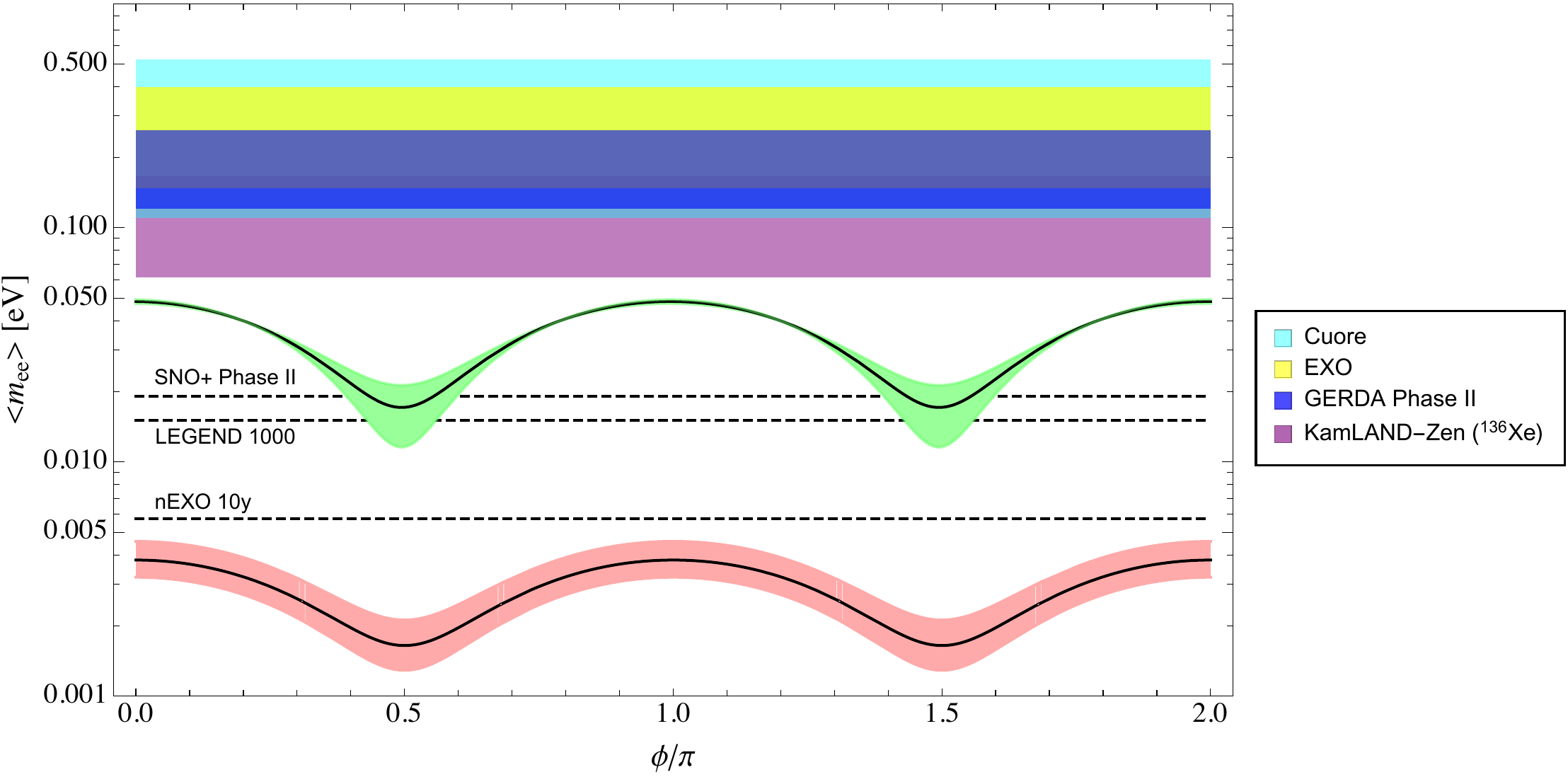} \vglue -.4cm 
\caption{\znbb decay amplitude in a generic three-neutrino scheme (left) 
  and in a two-massive-neutrino-scheme~\cite{Reig:2018ztc,Avila:2019hhv}, right.
  The vertical nearly-degenerate band is excluded by cosmology~\cite{Lattanzi:2020iik}.}
    \label{fig:dbd}
  \end{figure}

{\bf Absolute neutrino mass scale}~
Beta decay endpoint studies and cosmological observations probe the absolute neutrino mass, inaccessible to oscillations.
For example, the Katrin experiment has set an upper limit of 1.1 eV (at 90\% C.L.)~\cite{Aker:2019uuj}, which holds irrespective of whether neutrinos are Dirac or Majorana.
In the latter case there is also a neutrinoless \znbb variety of double beta decay mediated by the virtual Majorana neutrinos.
In Fig.~\ref{fig:dbd}, left panel from Ref.~\cite{Lattanzi:2020iik}, we display the \znbb amplitute versus the degeneracy parameter $\eta$, so that $\eta\simeq1$ corresponds to nearly degenerate neutrinos.
The normal-ordered neutrino region is indicated in the left blue band, while inverted ordering gives the upper-right branch.
Thanks to the Majorana phases~\cite{Schechter:1980gr}, the \znbb amplitude can vanish for normal ordering~\footnote{Note however that in many models there is a \znbb lower bound even for normal ordering, since the cancellation is prevented by the flavor structure of the leptonic weak interaction vertex~\cite{Dorame:2011eb,Dorame:2012zv,King:2013hj}. However this is model-dependent.}.
The vertical band is excluded by cosmological observations~\cite{Lattanzi:2020iik}, while the horizontal ones denote the current limit and the future expected sensitivities.
 
  Note that if one of the three neutrinos is massless or nearly so, as in the ``missing partner'' seesaw~\cite{Schechter:1980gr},
  no cancellation is possible, even for normal-ordering~\cite{Reig:2018ztc,Avila:2019hhv}.   
  The resulting regions correlate with the only free parameter available, i.e. the relative neutrino Majorana phase, as seen in Fig.~\ref{fig:dbd}, right panel.
  There is a lower bound, though it lies below detectability in upcoming experiments.
  In contrast, the lower bound found for inverse ordering lies higher than in the generic three-neutrino case, giving hope that
the coming experiments~\cite{Agostini:2019hzm} may not only prove the Majorana nature by the black-box theorem~\cite{Schechter:1981bd},
but also determine the relevant Majorana phase.\\[-.4cm]

{\bf Origin of neutrino mass}~
Even though the \sm lacks neutrino masses, these can arise from a unique dimension-five operator characterizing lepton number non-conservation~\cite{Weinberg:1979sa}. 
Seesaw completions open the way for a \textit{dynamical} understanding of small neutrino masses.
The mediators, ``right-handed'' singlet neutrinos in type-I, or the scalar triplet in type-II, were originally thought to lie at a high mass scale, associated to SO10 unification.
In addition to the standard vacuum expectation value (vev) responsible for electroweak breaking, there are new ``neutrinophilic'' vevs.
The most general seesaw description of neutrino mass generation is obtained by adopting the minimal \SM gauge group~\cite{Schechter:1980gr}, instead of left-right symmetric extensions.
The new vevs drive the spontaneous violation of lepton number, accompanied by a Goldstone boson, dubbed Majoron~\cite{Chikashige:1980ui,Schechter:1981cv}.
Such dynamical 3-2-1 seesaw mechanism can improve the consistency properties of the electroweak vacuum, such as stability~\cite{Bonilla:2015eha,Bonilla:2015kna,Mandal:2019ndp}.

In such general seesaw one can have any number of ``right-handed'' neutrino mediators, since they are gauge singlets.
As an example, one may consider a seesaw scheme with only two ``right'' neutrinos, implying that one of the ``left'' neutrinos remains massless~\cite{Schechter:1980gr}. 
Likewise, a ``missing partner'' seesaw scheme with just one singlet mediator leads to two massless neutrinos.
A ``dark sector'' can provide the other (solar) neutrino scale in a ``calculable'' way, as in the scoto-seesaw~\cite{Rojas:2018wym}.
The lower bound on \znbb decay shown in Fig.~\ref{fig:dbd} (right panel) holds in both missing-partner cases.

Conversely, one may add more ``right-'' than ``left-handed'' neutrinos, for example, one can have two isosinglets per family, sequentially.
By imposing lepton number conservation on such scheme one finds that light neutrinos are massless, exactly as in the Standard Model, as long as lepton number symmetry is exact.
In contrast to the \sm case, however, lepton flavor is violated, and similarly, CP symmetry.
This has two important implications. On the one hand, it elucidates the meaning of flavor and CP violation in the leptonic weak interaction, implying that such processes
need not be suppressed by the small neutrino masses, and hence can be large.
%%%%%
This construction also provides a template for building genuine ``low-scale'' realizations of the seesaw mechanism by adding a small lepton
number violation seed~\cite{Mohapatra:1986bd,Akhmedov:1995vm,Akhmedov:1995ip,Malinsky:2005bi}.
In contrast to the high-scale seesaw mechanism, there is a rich phenomenology in low-scale seesaw, including non-unitarity effects in neutrino propagation,
large \lfv rates and the possibility of having the neutrino mass mediators accessible to collider energies.
For a discussion and original references see Chapter 15 in~\cite{Valle:2015pba}.

Apart from inessential technicalities, the basic seesaw philosophy also works for the case of Dirac neutrinos.
One has both type-I and type-II Dirac seesaw realizations, and the small neutrino masses are also symmetry-protected.
Notice that there might be a deep reason for neutrinos to be Dirac-type. For example, this could signal the stability of dark matter~\cite{Chulia:2016ngi,CentellesChulia:2017koy}
or the existence of an underlying Peccei-Quinn symmetry~\cite{Dias:2020kbj}. \\[-.4cm]

{\bf Neutrinos and the flavor and dark matter problems}~
The \sm lacks an organizing principle in terms of which to understand the replication of fermion families, their masses and mixings.
  Here lies one of the deepest problems in our field.
  The discovery of oscillations have left an important legacy in particle physics, by establishing that (i) neutrino masses exist and (ii)
  leptons mix differenty from quarks, with two large mixing angles.
Understanding this from first principles constitutes a formidable challenge. 
In trying to account for the observed neutrino mixing pattern within UV-complete 3-2-1 family symmetry models one discovers, perhaps, one of the ingredients towards the complete theory, namely
the ``golden quark-lepton mass relation''~\cite{Morisi:2011pt}
\begin{equation}
\frac{m_\tau}{\sqrt{m_\mu m_e}}\approx \frac{m_b}{\sqrt{m_s m_d}}~.
\label{eq:golden}
\end{equation}
Such relation resembles b-tau unification, except for the lack of an underlying SU(5) gauge group. Moreover, it is a genuine flavor prediction involving all of the three families.
It emerges as a common prediction of many family-symmetry
models~\cite{King:2013hj,Morisi:2011pt,Morisi:2013eca,Bonilla:2014xla,Bonilla:2017ekt,Reig:2018ocz,deAnda:2019jxw,deAnda:2020pti,deAnda:2020ssl}.\\[-.4cm]

\begin{figure}[h]
    \centering \vglue -.2cm
\includegraphics[height=4.5cm,width=0.45\textwidth]{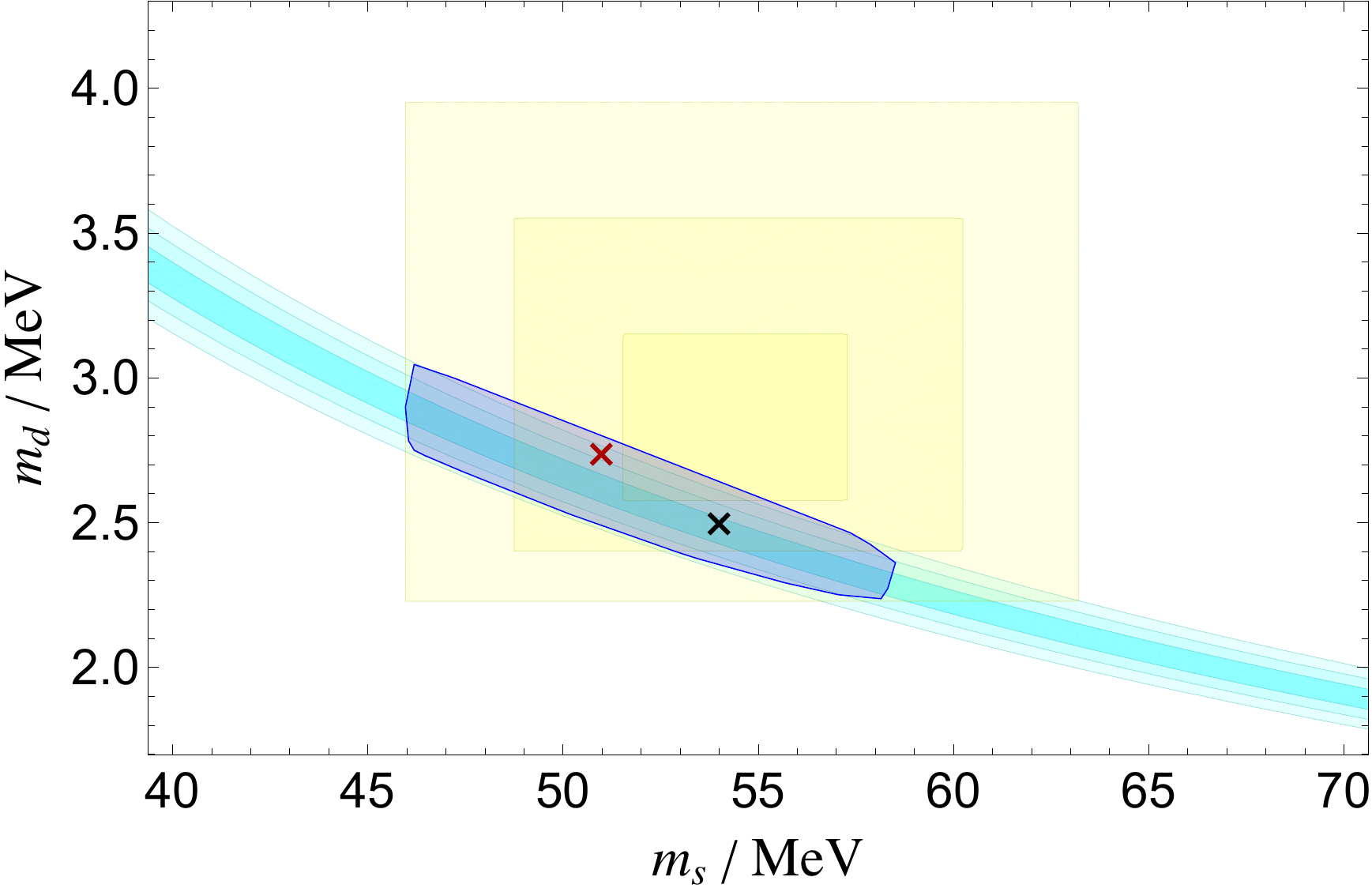}~~~~~%
\includegraphics[height=4.5cm,width=0.45\textwidth]{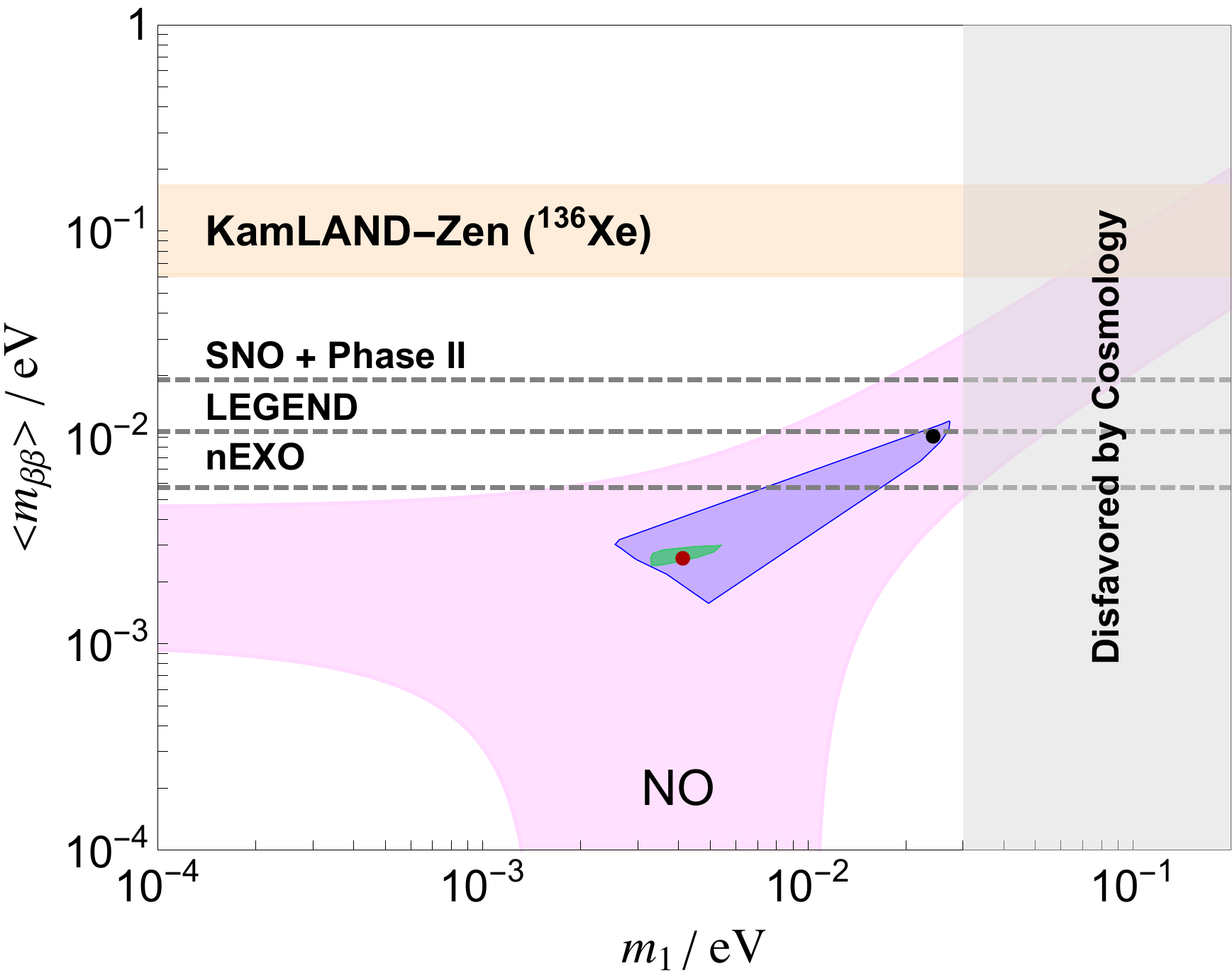}\vglue -.4cm 
\caption{Predictions of the ``golden'' quark-lepton unification formula (left), 
  and for the \znbb decay amplitude for \textbf{MI} (\textbf{MII}) scenarios, right. From~\cite{deAnda:2020pti}, see text.}
    \label{fig:flavor}
  \end{figure}

There are many ways to model flavor within UV-complete schemes, and I will refrain from entering into this arcane subject.
Instead, I will convey the radical idea that perhaps extra dimensions provide an adequate arena for pursuing a theory of flavor.
As an example, I discuss a recently proposed theory for fermion masses and mixings in which an \textbf{A4} family symmetry arises naturally from a six-dimensional spacetime after orbifold compactification.
As seen in the left panel in Fig.~\ref{fig:flavor} the flavor symmetry leads to the successful ``golden'' quark-lepton unification formula.
One sees the resulting prediction for the down- and strange-quark masses at the $M_Z$ scale.
  The cyan contours represent the 1,  2 and 3$\sigma$ allowed regions from the golden relation. 
  The yellow contours show the 1, 2 and 3$\sigma$ ranges of the measured quark masses at the $M_Z$ scale~\cite{Antusch:2013jca}.
  The blue region is the region consistent at 3$\sigma$ with the model's global flavor fit. The red (black) cross indicates the location of the best fit point for two model variants,
  \textbf{MI} (\textbf{MII}).

In the simplest ``constrained'' realization \textbf{MI} we assume that the Higgs vevs are real.
Assuming real Yukawa couplings, this implies that the only source of CP violation is numerically fixed.
This leads to a very strong predictivity for CP violation, including even the sign of the relevant CP phase.
CP violation of quarks and leptons has a common source and this leads to some tension in the global fit of the flavor observables, see~\cite{deAnda:2020pti} for details. 
The effective neutrinoless double-beta decay mass parameter is also sharply predicted, as seen in the right panel in Fig.~\ref{fig:flavor}.
  
  In the second model \textbf{MII} we relax the assumption that all the Higgs vevs are real, keeping real Yukawa couplings. 
  In this case the tension is resolved and the fit quality is excelent, with good predictions for \znbb and neutrino oscillations.
Upcoming experiments DUNE as well as LEGEND and nEXO offer good chances of exploring a substantial region of neutrino parameters.

 To conclude, let me comment that dark matter and neutrinos could be closely related. Out of the many ways to implement this connection, here I mention the so-called scotogenic approach,
 in which WIMP dark matter acts as the mediator of neutrino mass generation~\cite{Ma:2006km}, and which can also be neatly realized within the above orbifold scenario, see Ref.~\cite{deAnda:2020ssl}
 for details.
\tiny
\bibliographystyle{utphys}
%% commented on 31/12/19
\bibliography{bibliography}
\end{document}